\documentstyle[
aps
]{revtex}

\begin{document}

\draft

\title{Micellisation vs aggregation in dilute 
solutions of amphiphilic heteropolymers}

\author{
Edward G.~Timoshenko%
\setcounter{footnote}{0}\thanks{Author to 
whom correspondence should be addressed. 
Phone: +353-1-7162821,
Fax: +353-1-7162127.
Web page: http://darkstar.ucd.ie; 
E-mail: Edward.Timoshenko@ucd.ie},
Roman N. Basovsky
}
\address{
Theory and Computation Group,
Department of Chemistry, University College Dublin,
Belfield, Dublin 4, Ireland}

\author{Yuri A. Kuznetsov%
\thanks{
E-mail: Yuri.Kuznetsov@ucd.ie }}
\address{Centre for High Performance Computing Applications, 
University College Dublin, Belfield, Dublin 4, Ireland}

\date{\today}

\maketitle

\begin{abstract}
In recent experiments involving PNIPAM copolymers it has been observed that 
stable spherical nanoparticles are being formed by association 
of several chains in poor aqueous solution instead of aggregation. 
This type of mesoscopic structures called mesoglobules has an 
extremely monodispersed size distribution. 
Previously we have studied theoretically formation of  clusters 
consisting of several chains in dilute solutions of amphiphilic heteropolymers. 
We have seen that the mesoglobules often possess an essentially micellar 
structure.
In  the current work we argue that formation of mesoglobules is 
strongly sequence and concentration dependent. 
In particular, by means of lattice Monte Carlo 
simulation we consider structures formed by a number of  tri--block sequences 
and their certain mutations. For sequences consisting of two hydrophilic ends 
with a hydrophobic middle the size distribution of the resulting particles is 
rather narrow, i.e. mesoglobules are being formed. However, in the case of 
sequences with two hydrophobic ends and a hydrophilic middle a new type of 
structures seems to be prevalent. These consist of small mesoglobules 
interconnected by hydrophilic bridges. 
Clearly, on increasing the concentration 
these networked structures would play an important role resulting in a rapid 
onset of a gel--like behaviour.\\
$\left.\right.$
\end{abstract}

\section{Introduction}\label{sec:intro}

In recent years there were numerous works devoted to the behaviour
of water soluble polymers near the lower critical solubility
temperature (LCST). Typical systems include
poly-N-isopropylacrylamide (PNIPAM) homopolymer and
block copolymers \cite{Schild92,Fuj89} of the 
poly(ethylene oxide)-poly(propylene oxide)-poly(ethylene oxide),
or PEO-PPO systems briefly \cite{peoppo1,pluronicExp}.
The latter copolymers are widely used in pharmaceutical, agricultural
and food industries due to their low toxicity and surfactant
characteristics (EO is hydrophilic and PO is hydrophobic).
These are commercially available as Pluronic or Synperonic polymers
and they have been quite well studied experimentally 
\cite{pluronicExp,peoppo1}.
Speaking of tri--block copolymers in particular 
there are two types which have rather distinct properties. 
First, a PEO-PPO-PEO polymer with the hydrophilic
ends and hydrophobic middle is the standard Pluronic which forms stable
micelles.
Second, a polymer with the inverse structure would tend to bridge
hydrophobic end `stickers' by hydrophilic bridges.
These are so--called telechelic associating polymers
\cite{Telechelic}, which form telechelic
gels possessing nontrivial rheological properties at higher concentrations.

The phase behaviour of PNIPAM homopolymer has been
well investigated \cite{Schild92,Fuj89}.
PNIPAM serves as an important model polymer
due to its convenient LCST in water which is at approximately $32^o$ C, i.e. 
near the room temperature.
One of the important applications of this polymer is based on the
use of temperature and pH responsive PNIPAM gels, which exhibit
dramatic conformational changes such as swelling upon small changes
of the external conditions such as temperature or pH.
One of popular modifications is achieved by grafting
PNIPAM main chain with PEO side--groups. After collapse of the backbone chain
the PEO side chains solubilise it in water.

Polymers which form micellar structures generally are of great interest for
biological and pharmaceutical applications because they may be
used in a drug delivery process due to their ability to solubilise
hydrophobic compounds and due to an improved stability as compared to
low molecular weight systems.
At higher concentrations these systems
display a very rich variety of phases and corresponding morphologies
\cite{peoppo1}.
One is particularly interested in the properties of dilute solutions as
they are determined by the fundamental intra--molecular interactions.
Experiments in this area \cite{Ricka91,Chu95,Gorelov94}
are quite difficult and often obscured
by inter--polymer aggregation phenomena.

Perhaps one of the most striking recent experimental observations was
the formation of rather size monodispersed nanoparticles \cite{Gorelov94}. 
This phenomenon has been recently observed by a number of Groups worldwide:
R.~Pelton {\em et al} \cite{Pelton} at McMaster University, by Chi Wu 
{\em et al} \cite{Wu} in Hong Kong,
A.V. Gorelov {\em at al} \cite{Gorelov94} at UCD, and some indications
in favour of these structures have also been seen in Lund University
\cite{Schillen}.
The appearance of spherical particles in
dilute solutions of poly(N-isopropylacrylamide) has been clearly observed by
dynamic light scattering (DLS). 
Significantly, these particles have a relatively narrow size
distribution and remain stable for many days. 
Thus, the size of the particles is thermodynamically controlled and
it increases with increasing  
polymer concentration and increasing strength of inter--monomer attraction. 
Electron microscopy further confirms that these spherical particles 
have mean size and distribution in agreement with DLS.

We have recently been able to explain the stability of the mesoglobules
based on the extended version of the Gaussian variational theory
\cite{ConfTra} as well as to determine the histograms of their
size distributions via lattice Monte Carlo simulations in Ref. \cite{Clusters}.
The existence of such mesoscopic structures is  maintained by a delicate 
balance of the energetic  and entropic terms under the connectivity constraints.
In recent paper Ref. \cite{ClusEPL} we have examined some 
more complex heteropolymer sequences, such as tri--block copolymers. 
The main conclusion from such a study was that in the case of tri--blocks
with outer hydrophilic blocks mesoglobules are thermodynamically
stable in a narrow region of the phase diagram in accordance with
the expectations from our theory. The case of tri--blocks with the outer
hydrophobic blocks, however, generally does not have this property
since the hydrophobic end `stickers' can attach to similar blocks
from other macromolecular clusters which thus are becoming connected via
extended hydrophilic bridges. This picture is quite familiar from theory
of associating polymers. Indeed, telechelic polymers \cite{Telechelic}
at higher concentrations would become highly inter--connected
producing a physical gel.

The purpouse of the current work is to consider a wider range of
heteropolymer sequences in order to examine the competition of
aggregation vs micellesation when the latter is accompanied by formation
of mesoglobules.
First, we shall proceed from the tri--block copolymers and apply
certain `mutations' obtained by permutation of some monomers to examine
the stability of the clusters and how the mean size and dispersion
of cluster sizes would be affected. 
Second, we shall examine the effect of decreasing the chain length
with a simultaneous increase of the concentration at a constant 
total number of monomers in the system.

\section{Results}
\label{sec:results}

We adopt the Metropolis technique in the lattice model of
Ref. \cite{Clusters}.
This model, apart from the connectivity and excluded volume constraints,
includes short--ranged pair--wise interactions between
lattice sites.
The system is completely characterised by three Flory interaction
parameters, $\chi_{aa}$, $\chi_{ab}$ and $\chi_{bb}$,
along with $N$, $M$ and linear lattice size $L$.
In addition to local monomer moves \cite{Clusters}, we
include translational moves representing diffusion of chains.
The latter moves are applied to all clusters
of chains with a probability inversely proportional to
the number of monomers within (Stokes law).

To address the question about the size polydispersity of
the mesoglobules we have obtained a large ensemble
of independent equilibrium states in the region of the phase
diagram where mesoglobules can be formed. Then by using the histogram
technique we obtain the
probability densities of the number of chains in a cluster 
(aggregation number), $M_{cl}$,
and of the radius of gyration of a cluster (size), $R_{cl}$
for different considered sequences.

In Figs.~\ref{fig:hysto20} we present these distributions for 
the OUTER-H (s1) and OUTER-P (s2) tri--blocks, as well as for
their mutations s3 and s4 respectively, which are obtained by
point--like permutations of two monomers near the points where
a and b blocks join each other.
The s2 sequence has a single high and narrow peak in both
distributions, which corresponds to a well defined mean size of the
well monodispersed mesoglobules. Sequence s1, however, has two
peaks, which are also broader and less pronounced, as well as
a high population of the macro--aggregate (cluster of $M_{cl}=M=20$).
This is due to bridging of hydrophobic clusters by the hydrophilic
middle blocks.
Mutated sequences s3 and s4 have quite close distributions for
the aggregation numbers $P(M_{cl})$ to
their respective tri--blocks s1 and s2, demonstrating the stability
of these two types of structures with respect to slight perturbations
in the primary sequence. However, in the size distributions $P(R_{cl})$
the locations of the peaks have shifted somewhat to the left.
Therefore, both the mesoglobules and the aggregates for the mutated
sequences have become somewhat smaller compared to those of the tri--blocks.
This is natural since the outstretched hydrophilic tails or loops
have become somewhat shorter.

In Figs.~\ref{fig:hysto40} we exhibit the two distributions for
the system composed of twice the number of chains $M=40$ with
half--long polymers $N=12$ in the same simulation box as compared
to Figs.~\ref{fig:hysto20}.
For OUTER-P sequence s2 there is still a single peak corresponding
to mesoglobules. 
A typical snapshot for these is presented in Fig.~\ref{fig:snaps}a,
where one can see four micelles of essentially equal size.
Compared to the twice longer sequence
in Figs.~\ref{fig:hysto20} the size of the mesoglobules has become
smaller, although the aggregation numbers have increased in approximately
$1.5$ times. The distributions for s1 sequence still have two peaks,
the second corresponding to a population of the macro--aggregate. 
Thus, the nearly equal sized
clusters are present there also, but in a small fraction, but
many members of the ensemble have joined into dimers, trimers and
so on formed from spherical clusters as shown in Fig.~\ref{fig:snaps}b.
Interestingly, both s1 and s2 have rather small populations of
the globules of single chains, which is due to the short length
of the chains.

Distributions for the mutated sequences s3 and s4, however,
are quite different from those of s1 and s2. Namely,
for s4 the mesoglobules have become considerably larger
and somewhat less pronounced due to an increased population of
single chain globules compared to s2. The shape of distributions for
s3 compared to s1 have undergone quite dramatic changes.
First, the population of single chain globules have increased considerably.
Second, the mean size of large clusters have increased and these have
become more monodispersed. Third, the population of
macro--aggregate has decreased. A typical snapshot for this
situation is presented in Fig.~\ref{fig:snaps}c, where one
can see a macro--aggregate of $M_{cl}=36$ chains coexisting with four
single chain globules.
Quite remarkably, the mutations of s1 has helped to stabilise
average sized clusters leading to more size monodispersed `mesoglobules'.


\section{Conclusion}

In this paper we have studied the effect of point--like mutations in
tri--block copolymers on aggregates and mesoglobules in dilute solutions.
For sufficiently long chains the effect of small mutations is rather weak
and corresponding structures remain stable.
However, larger--scale mutations may produce stronger influence.

We then looked at solutions of twice the number of half--long chains
for similar sequences. In this case, however, the shortness of the chains
has added certain populations of single chain globules for all sequences.
The main effect of mutations was in dramatically increasing
the populations of single chain globules as compared to the
original tri--blocks.
Perhaps the most unexpected result of mutations was in pronounced increase
of the mesoglobules size (both in terms of radius and aggregation number)
for mutated sequences compared to the tri--blocks. In Ref. \cite{Clusters}
we have argued that the size of the mesoglobules is determined by
a characteristic scale of micro--phase separation, which in turn is
related to the lengths of the blocks. Thus, mesoglobules of  shorter
blocks should be smaller according to this line of arguing.
This was so for sequences of length $N=24$ in Figs.~\ref{fig:hysto20}
because mutations effectively decrease the lengths of blocks.
However, in Fig.~\ref{fig:hysto40} the mutations have produced
the opposite effect of pronounced magnitude. This paradoxical result
reflects the complexity of the competitions of different interactions
and the entropy and it does not have a simple explanation as yet.

The results of this work are important for understanding the role
of various factors and imperfections in the design of polymers capable
of producing stable monodispersed mesoglobules.
Mesoglobules with well controlled size distribution
can find a range of applications in pharmaceutical,
biotechnological and cosmetic industries.
Moreover, these studies may also help
understanding the problems of competition aggregation vs folding
in protein solutions and, possibly,
self--organisation of the quaternary structures in
multimeric proteins.

\acknowledgments

The authors acknowledge interesting discussions with Professor
F.~Ganazzoli, Professor H.~Orland and Professor T.~Garel.
This work was supported by grant SC/99/186 from Enterprise Ireland.





\begin{figure}
\caption{
Probability densities of the number of chains in a cluster 
(aggregation number), $M_{cl}$,
(Fig.~a) and of the radius of gyration of a cluster (size), $R_{cl}$, 
(Fig.~b) for tri--blocks (thick lines) and mutated tri--blocks (thin lines) 
sequences.
These results have been obtained by analyzing data for
the ensemble size $Q=1000$ for each sequence.
Here the equilibration time was
$1.92 \cdot 10^9$ of attempted Monte Carlo moves and other
parameters were: linear lattice size $L=60$, degree of polymerisation $N=24$,
number of chains $M=20$, and the Flory interaction parameters 
$\chi_{aa}=1$, $\chi_{ab}=0.4$ and $\chi_{bb}=-0.2$.
}\label{fig:hysto20}
\end{figure}

\begin{figure}
\caption{
Probability densities of the number of chains in a cluster 
(aggregation number), $M_{cl}$,
(Fig.~a), and the radius of gyration of a cluster (size), $R_{cl}$ (Fig.~b)
for short tri--blocks (thick lines) and mutated tri--blocks (thin lines)
sequences. Here $N=12$ and $M=40$, while
other parameters are as in Fig.~\ref{fig:hysto20}.
}\label{fig:hysto40}
\end{figure}

\begin{figure}
\caption{
Snapshots of typical conformations from simulation for different sequences.
Figs.~a, b and c correspond to sequences $b_3 a_6 b_3$,
$a_3 b_6 a_3$ and $a_2 b a b_4 a b a_2$ respectively
(these were denoted as $s2$,
$s1$ and $s3$ respectively in Fig.~\ref{fig:hysto40}).
Here black (white) circles correspond to hydrophobic (hydrophilic)
monomers, while
other parameters are as in Fig.~\ref{fig:hysto40}.
}\label{fig:snaps}
\end{figure}


\begin{references}

\bibitem{Schild92}
H.G. Schild,  
Prog. Polym. Sci. 17 (1992) 163.

\bibitem{Fuj89}
K. Kubota, S. Fujishige, I. Ando,
Polym. J. 22 (1990) 15.

\bibitem{peoppo1}
Y.M. Lam, N. Grigorieff, G. Goldbeck-Wood,
Phys. Chem. Chem. Phys. 1  (1999) 3331.

\bibitem{pluronicExp}
G. Wanka, H. Hoffmann, U. Ulbricht,
Macromolecules 27  (1994) 4145;
P. Alexandridis, T.A. Hatton, 
Colloids Surfaces A 96 (1995) 1.

\bibitem{Telechelic}
A.N. Semenov, J.-F. Joanny, A. Khokhlov,
Macromolecules 28 (1995) 1066.

\bibitem{Ricka91}
M. Meewes, J. Ri\v{c}ka, M. de Silva, R. Nyffenegger, T. Binkert,
Macromolecules 24 (1991) 5811.

\bibitem{Chu95}
B. Chu, Q.C. Ying, A.Yu. Grosberg,
Macromolecules 28 (1995) 180.

\bibitem{Gorelov94}
A.V. Gorelov, A. Du Chesne, K.A. Dawson,
Physica A 240 (1997) 443;
A.V. Gorelov, L.N. Vasileva, A. Du Chesne et al.
Nuovo Cimento D 16 (1994) 711.

\bibitem{Pelton}
K. Chan, R. Pelton, J. Zhang,
Lanmuir 15 (1999) 4018;
X. Wu, R.H. Pelton, A.E. Hamielec, et al,
Colloid. Polym Sci. 272 (1994) 467.

\bibitem{Wu}
C. Wu, J. Gao, M. Li, et al,
Macromol. Symp. 150 (2000) 219;
C. Wu, A.Z. Niu, L.M. Leung, et al,
J. Am. Chem. Soc. 121 (1999) 1954;
X.~Qiu, C.~M.~S.~Kwan, and C.~Wu, Macromolecules  30  (1997) 6090.

\bibitem{Schillen}
K. Schill\'en, K. Bryskhe, Yu.S. Mel'nikova,
Macromolecules 32 (1999) 6885.

\bibitem{ConfTra}
E.G. Timoshenko, Yu.A. Kuznetsov, K.A. Dawson,
Phys. Rev. E 57 (1998) 6801.

\bibitem{Clusters}
E.G. Timoshenko, Yu.A. Kuznetsov, 
J. Chem. Phys., 112 (2000) 8163.

\bibitem{ClusEPL}
E.G. Timoshenko, Yu.A. Kuznetsov,
Europhys. Lett., in press  (2000).


\end{references}
\end{document}